\documentclass[final,1p,times]{elsarticle} 
\usepackage{graphicx}
\usepackage{amssymb} 
\usepackage{amsthm}
\usepackage{amsmath}
\usepackage{lineno}
\usepackage{epsfig}
\usepackage{wrapfig}
\usepackage{caption}
\usepackage{subcaption}
\newcommand{\avbT}         {\ensuremath{\left< \beta_{\rm T}\right>}}
\newcommand{\Tfo}          {\ensuremath{{T}_{kin}}}
\newcommand{\Tch}          {\ensuremath{{T}_{ch}}}
\newcommand{\snn}          {\ensuremath{\sqrt{s_{\rm NN}}}}
\newcommand{\gevc}         {\ensuremath{{\rm GeV}/c}}
\newcommand{\pt}           {\ensuremath{p_{\rm T}}}
\newcommand{\PbPb}         {\mbox{Pb--Pb}}
\newcommand{\AuAu}         {\mbox{Au--Au}}
\newcommand{\dedx}         {\ensuremath{\mathrm{d}E/\mathrm{d}x}}
\newcommand{\ITS}          {\rm{ITS}}
\newcommand{\TPC}          {\rm{TPC}}
\newcommand{\TOF}          {\rm{TOF}}
\newcommand{\VZERO}        {\rm{VZERO}}
\newcommand{\dcaxy}        {\ensuremath{{\rm DCA}_{xy}}}
\newcommand{\muB}          {\ensuremath{\mu_{B}}}

\newcommand{\dndy}         {d$N$/d$y$}

\journal{Nuclear Physics A} 

\begin{document}

\begin{frontmatter} 

% Your Title - please insert
\title{Identified charged hadron production in \PbPb\ collisions at the LHC with the ALICE Experiment}

%% Single author (and collaboration) - please insert
\author{Leonardo Milano (for the ALICE\fnref{col1} Collaboration)}
\fntext[col1] {A list of members of the ALICE Collaboration and acknowledgements can be found at the end of this issue.}
\address{Dipartimento di Fisica dell'Universit\'a di Torino \& Istituto Nazionale di Fisica Nucleare - INFN}

%% Multiple authors
%\author[auth2]{Marcus Junius Brutus}
%\address[auth1]{Somewhere, Rome}
%\address[auth2]{Somewhere else, Rome}

\begin{abstract} 
Identified particle spectra represent a crucial tool to understand the behavior of the matter created in high-energy heavy-ion collisions.
The transverse momentum \pt\ distributions of identified hadrons contain informations about the transverse expansion of the system and constrain the freezeout properties of the matter created.
The ALICE experiment has good particle identification performance over a broad \pt\ range.
%Particles are identified using the energy loss signal in the Inner Tracking System and Time Projection Chamber detectors, complemented with the information from the Time of Flight detector to identify hadrons up to \pt~$\sim$~4.5 \gevc.\\
In this contribution the results for identified pions, kaons and protons in heavy-ion collisions at 2.76 TeV center-of-mass energy are presented. These results are compared with other identified particle measurements obtained by previous experiments, and discussed in terms of the thermal and hydrodynamic pictures.
The status of extensions of this analysis, with the study of identified particles as a function of event-by-event flow in \PbPb\ collisions, is also discussed.
\end{abstract} 

\end{frontmatter} % do not change

%% linenumbers are useful for reviewing process
%\linenumbers

%\section{Introduction}
%\label{sec:Intro}
The ALICE experiment has unique particle identification (PID) capabilities. The combination of different detectors which use different PID techniques allows identification over a broad \pt\ range \cite{KarelQM2012}. The results presented here are obtained using the \PbPb\ data at \snn\ = 2.76 TeV collected at the Large Hadron Collider (LHC) during the fall 2010. The central tracking and (PID) detectors cover the pseudorapidity window $|\eta| \leq$ 0.9.
%Tracks are reconstructed using the tracking detectors in the central barrel. A specific tracking algorithm which uses only the Inner Tracking System (ITS-standalone tracking) can extend the low \pt~reach and recover tracks missed by the Time Projection Chamber (\TPC).
Particles are identified using \dedx~signal in the silicon Inner Tracking System (\ITS) and in the Time Projection Chamber (\TPC) and the information from the Time Of Flight (\TOF)~detector. 
A pair of forward scintillator hodoscopes, the \VZERO\ detectors (2.8 $< \eta <$ 5.1 and -3.7 $< \eta <$ -1.7), is used for triggering \cite{ALICECentrality}.
The centrality of the collision can be estimated using the signal in the \VZERO~detector, the reconstructed multiplicity in the central barrel, or other forward detectors \cite{ALICECentrality}.

\section{\pt\ distribution of identified hadrons}
\begin{figure}[htbp]
  \centering
 \begin{subfigure}[b]{0.45\textwidth}
         \includegraphics[width=0.9\textwidth]{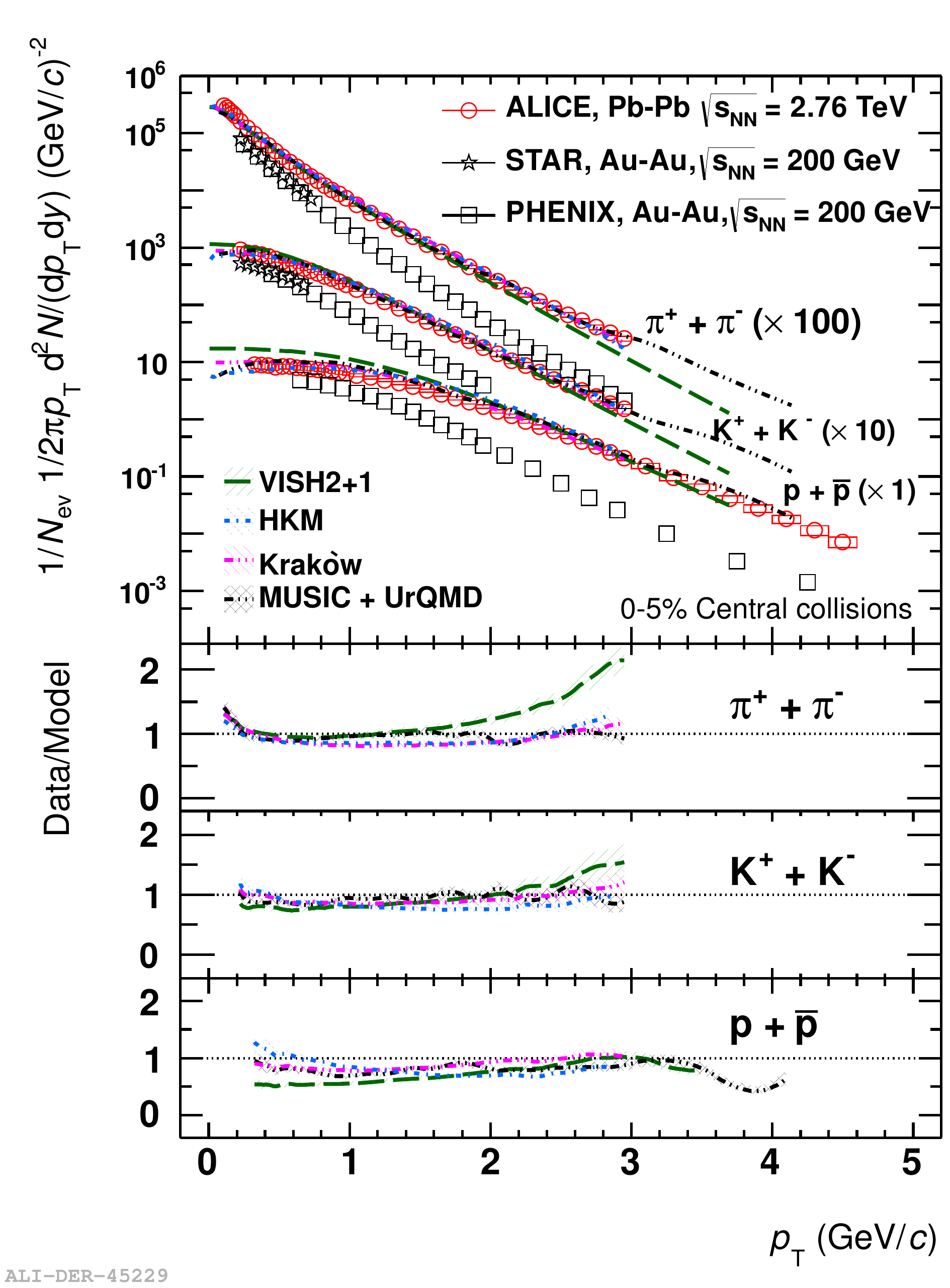}
 \end{subfigure}
\begin{subfigure}[b]{0.54\textwidth}
 \includegraphics[width=\textwidth]{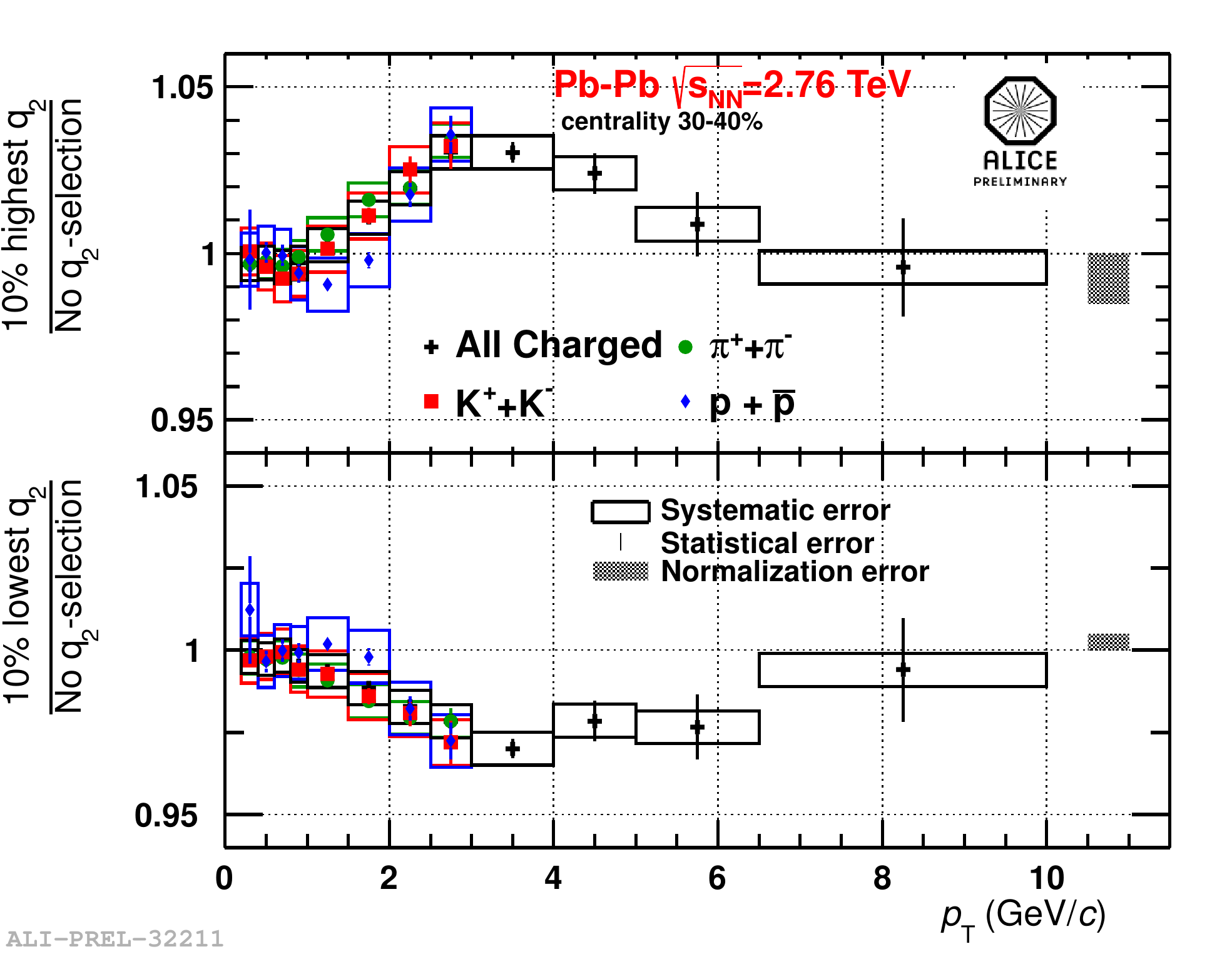}
 \end{subfigure}
\caption{Left: transverse momentum distributions  of the sum of positive and negative particles (box: systematic errors; statistical errors smaller than the symbol for most data points) compared to RHIC data and hydrodynamic models. Right: ratio between the raw spectra in the sample with 10\% highest (lowest) $q_2$ and the unbiased sample.}
\label{fig:spectra-central}
\end{figure}

%\subsection{Identified particle spectra in central (0-5\%) \PbPb\ collisions at \snn\ = 2.76 TeV}
\label{sec:PID}
The \pt\ distribution of hadrons contains the information about the collective expansion of the fireball (radial flow) and the temperature of kinetic freezeout (\Tfo). 
The ALICE measurement of identified particle spectra in central (0-5\%) \PbPb\ collisions at \snn\ = 2.76 TeV is represented by the red symbols in Figure \ref{fig:spectra-central}, left (from \cite{ALICECentralPbPbSpectra}). The \pt\ distributions of positive and negative particles are found to be compatible within errors, for this reason results for summed charge states are presented.
%The measurement spans the \pt\ range from $\sim$~0.1 \gevc~up to $\sim$~4.5 \gevc.
%\begin{wrapfigure}{r}{0.5\textwidth}
%\begin{figure}[htbp]
%  \centering
%  \includegraphics[width=0.45\textwidth]{SpecraFigurePRLMUSIC}
%  \includegraphics[width=0.45\textwidth]{2012-Aug-07-SpecravsEbEFlow_ratio}
%\caption{Left: transverse momentum distributions  of the sum of positive and negative particles (box: systematic %errors; statistical errors smaller than the symbol for most data points) compared to RHIC data and hydrodynamic models. Right: ratio between the raw spectra in the sample with 10\% highest (lowest) $q_2$ and the unbiased sample.}
%  \label{fig:spectra-central}
%\end{wrapfigure}
Hadron spectra are reported for primary particles, defined as prompt particles produced in the collision, including decay products, except those from weak decays of strange particles.
The fraction of primaries in the track sample at a given \pt\ bin was estimated from data by fitting the \dcaxy\ distribution with three Monte Carlo templates: primary particles, secondaries from material and secondaries from weak decays \cite{AlexQM2011}.
A detailed description of the systematic error and PID procedure can be found in \cite{ALICECentralPbPbSpectra}. 
Spectra measured at the LHC are compared with RHIC results for \AuAu\ collisions at \snn\ = 200 GeV (black-empty markers). The spectral shape is significantly harder at the LHC with respect to RHIC. The kinetic freezeout parameters \avbT~and \Tfo\ can be extracted from a simultaneous blast wave fit to the $\pi$, K and p spectra. The \pt\ ranges used in the fit are 0.5-1~\gevc, 0.2-1.5~\gevc, 0.3-3~\gevc~for $\pi$, K and p. Data are well described by the blast wave function with \avbT\ = 0.65 $\pm$ 0.02 and \Tfo\ = 95 $\pm$ 10 MeV. It should be noted that \Tfo\ is sensitive to the pion fit range (due to large contribution from resonances) while \avbT~does not strongly depend on the \pt\ range used in the fit. A similar fit to central \AuAu\ collisions at \snn\ = 200 GeV was performed in \cite{STARSpectra2008}: the \avbT\ is $\sim$~10\% larger at the LHC with respect to RHIC and \Tfo\ is compatible within errors.
%Different hydrodynamic models are reported in Figure \ref{fig:spectra-central} .
%In the \pt\ range below 1.5 \gevc~a pure viscous hydrodynamic model (VISH2+1 \cite{VISH2+1}) describes quite well the $\pi$ and K spectra, but it misses the protons, both in shape and absolute yields. This discrepancy may be due to the lack of an explicit description of the hadronic phase in the model. This is supported by the comparison with the HKM \cite{HKM} model, in which the hadronic phase of the fireball evolution is described by UrQMD \cite{UrQMD}. The third model is the Krak\`ow model \cite{Krakov}, which introduces non-equilibrium corrections due to viscosity at the transition from the hydrodynamic description to particles, which change the effective chemical freezeout temperature \Tch, leading to a good agreement with the data. 
%During this conference a new event-by-event 3-D viscous hydrodynamic model (MUSIC) coupled with UrQMD afterburner was presented \cite{MUSIC}. The agreement with the data is good (the disagreement with protons at high \pt\ can be explained in terms of contribution from jets and mini-jets which is not included in the model).
%The comparison with theory seems to support the hydrodynamic interpretation of \pt\ spectra at the LHC.
The comparison with predictions (see \cite{ALICECentralPbPbSpectra} for details) seems to suggest that hydrodynamic models with a refined late fireball description are able to reproduce the measured \pt\ spectra at the LHC.

%\subsection{Identified Particle spectra as a function of event-by-event elliptic flow in semi-central (30-40\%) \PbPb~collisions at \snn\ = 2.76 TeV}
In order to further investigate the hydrodynamic behaviour of the hadron production two of the main features of hydrodynamics were correlated: \pt\ distribution of hadrons and elliptic flow.
%The integrated elliptic flow at the LHC was found to be $\sim$ 30\% larger with respect to RHIC value \cite{ALICEv2}. If one looks at the elliptic flow on an event-by-event basis this increase can be much larger. This is due to the fact that 
For a given centrality the eccentricity of the collision (related with the initial geometry) fluctuates.
A strategy to select events based on the geometry of the overlapping region (so called ``event shape engineering") was presented for the first time during this conference in \cite{SergeyQM2012,AlexandruQM2012}. One way to do this is using the \VZERO\ detector to calculate the flow vector $\vec{Q_2}$ on an event-by-event basis, a 2D vector with components $Q_{2,x}=\sum^{}_{i}w_i\cos(2\phi_i)$ and $Q_{2,y}=\sum^{}_{i}w_i\sin(2\phi_i)$,
where the sum $i$~runs over all the channels of the \VZERO\ detector, $w_i$ is the multiplicity of channel $i$~and $\phi_i$~is the angle of channel $i$. The module of the $\vec{Q_2}$ vector is normalized by the multiplicity $M$ in the \VZERO: $q_{2}=|Q_{2}|/\sqrt{M}$.
%Based on the $q_{2}$ distribution it is possible to select the 10\% highest (lowest) elliptic flow events.
%In order to get rid of the multiplicity bias (since $v_2$ is not constant as a function of centrality in 30-40\% \cite{ALICEv2}) the centrality is selected using tracks in the central barrel and the bin 30-40\% is obtained as the sum of 10 bins 1\% wide.
%In order to be sure that the large $q_2$ is not due to an increased jet contribution a raw jet cone algorithm is used to measure the jet contamination in the selected samples. A cone of radius $r=\sqrt{\Delta{\eta}^2 + \Delta{\phi}^2} \leq 0.3$ is defined around a seed particle with \pt\ $\geq$ 5 \gevc. The raw jet energy is estimated subtracting the background from the energy measured in the cone. The raw jet spectra in the samples selected according to the $q_2$ distribution and the unbiased sample are found to be compatible within errors.
In \cite{SergeyQM2012,AlexandruQM2012} it has been shown that the sample with large (small) $q_2$ shows a significantly larger (smaller) $v_2$ with respect to the unbiased sample and that the non-flow contributions are negligible. Systematic checks show that this selection does not introduce trivial biases related with multiplicity shift or jet contribution.
%The tracking and PID efficiency are found to be independent from the $q_2$ selection applied.
%\begin{figure}[tb]
%  \centering
%  \includegraphics[width=0.5\textwidth]{2012-Aug-07-SpecravsEbEFlow_ratio}
%\caption{Top (bottom): ratio between the raw spectra in the sample with 10\% highest (lowest) $q_2$ and the unbiased sample.}
%  \label{fig:Ratio-Q}
%\vspace{-10pt}
%\end{figure}
%Figure \ref{fig:Ratio-Q} shows on top (bottom) the ratio between the raw spectra in the sample with 10\% highest (lowest) $q_2$ and the unbiased spectra for unidentified charged hadrons and for identified $\pi$, K and p.
A modification  of the \pt\ spectrum in semi-central (30-40\%) \PbPb~collisions at \snn\ = 2.76 TeV in the intermediate \pt\ region (from $\sim$ 1 up to $\sim$ 5 \gevc) is observed, when events are selected according to the event shape engineering (Figure \ref{fig:spectra-central}, right): higher (lower) $v_2$ means harder (softer) spectra. This modification vanishes at high \pt~supporting a correlation related with hydrodynamics rather than with hard processes. A hint of mass ordering can be observed in the region between $\sim$ 1 up to $\sim$ 3 \gevc. A more detailed study (including comparison with models and hydrodynamic fit of particle spectra) will allow to better understand the observed correlation between $v_2$ and radial flow.

\section{Thermal production of hadrons at the LHC}
\label{sec:Thermal}
\begin{figure}[htbp]
  \centering
 \begin{subfigure}[b]{0.45\textwidth}
   \includegraphics[width=0.95\textwidth]{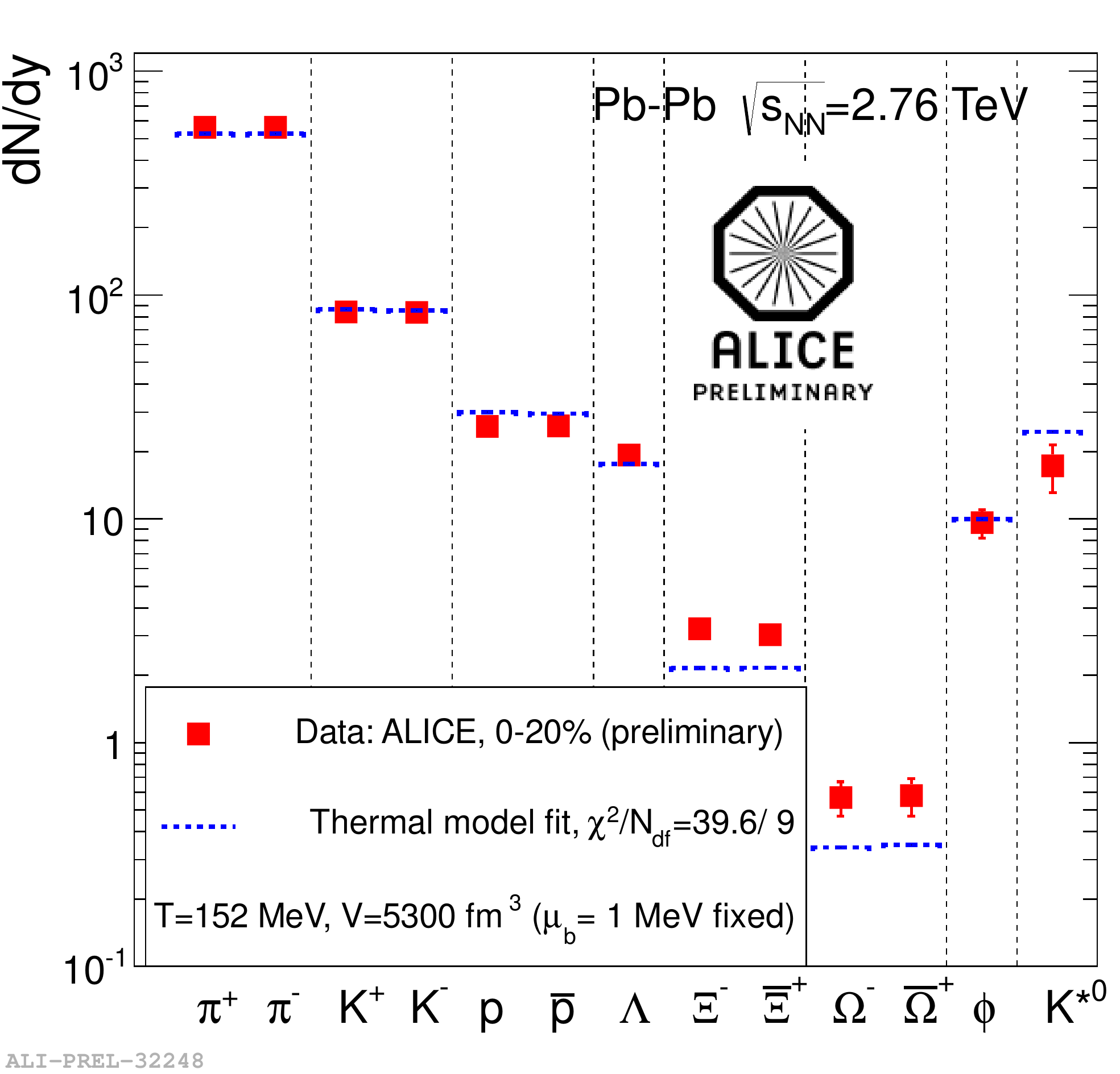}
 \end{subfigure}
\begin{subfigure}[b]{0.45\textwidth}
 \includegraphics[width=0.95\textwidth]{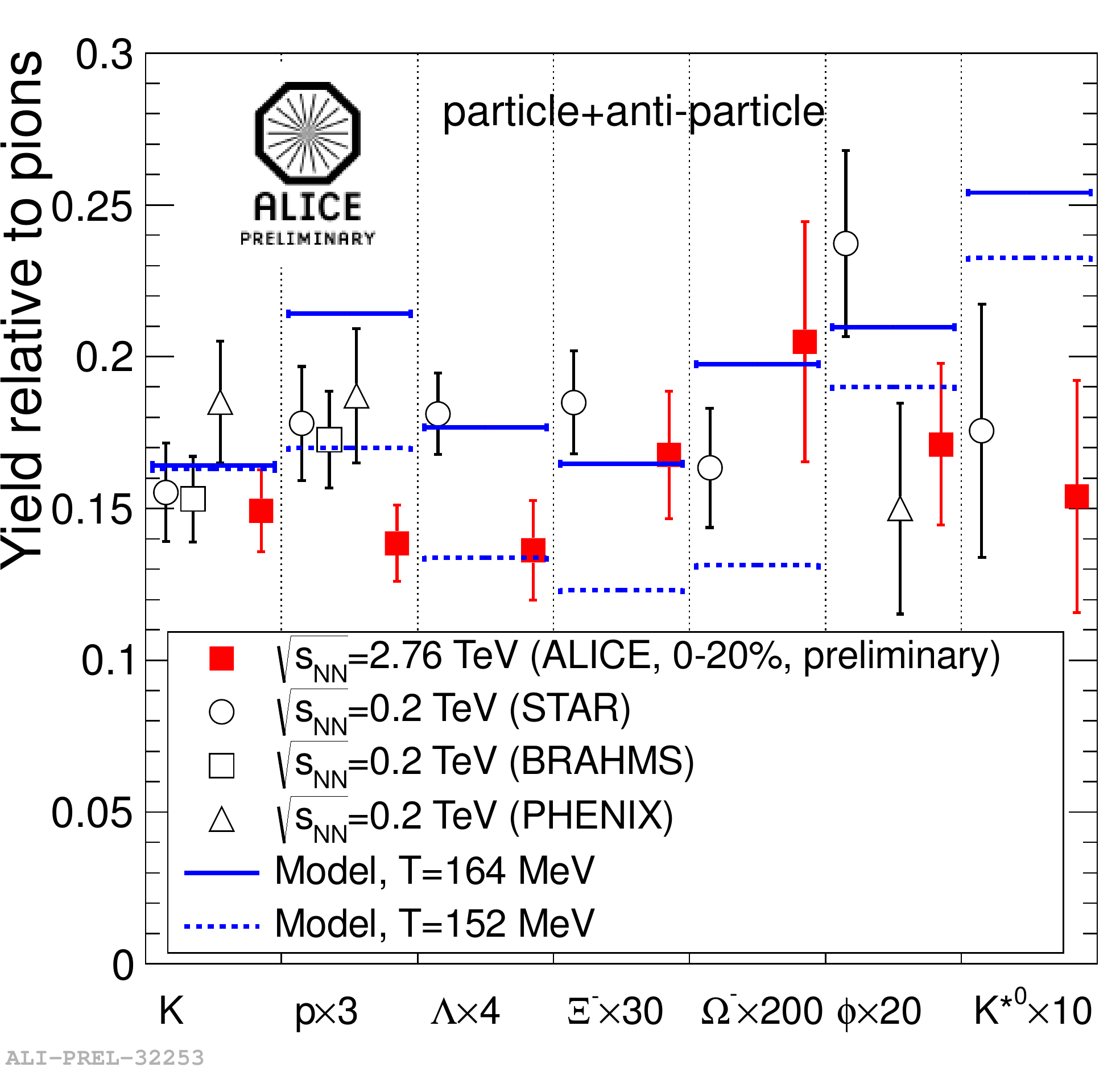}
 \end{subfigure}
\caption{Left: integrated yields at mid-rapidity \dndy~in central (0-20\%) \PbPb~collisions at \snn=2.76 TeV with results from thermal fit. Right: integrated yields at mid-rapidity \dndy~in central (0-20\%) \PbPb~collisions relative to pions with RHIC comparison and thermal model predictions.}
\label{fig:yields}
\end{figure}

%The \pt\ spectra of identified hadrons have been fitted individually using a blast wave function, in order to extract the integrated yield at midrapidity ($|$y$|<$ 0.5). Thanks to the good tracking and PID capability of the ALICE detector at low \pt\ the fraction of extrapolated yield is small (7\%, 6\%, and 4\% for $\pi$, K and p).
The thermal description of hadron production was found to be successful over a broad range of energies (from \snn\ = 2 GeV to \snn\ = 200 GeV \cite{PBM2009,Cleymans1998}).
There are only these three parameters which govern the thermal model: the chemical freezeout temperature \Tch, the baryon-chemical potential \muB\ and the volume $V$.
%\Tch\ is constant above SPS energies and \muB~decreases with increasing energy. 
In order to extract the parameters \Tch, \muB~and $V$~a thermal fit \cite{PBM2009} of integrated yields at midrapidity \dndy\ in central (0-20\%) \PbPb\ collisions was performed. It is reported in Figure \ref{fig:yields} (left). Results from strange and multi-strange particle analyses \cite{SubhashQM2012} are also included in the fit. The antibaryon over baryon ratios suggest a vanishing baryon-chemical potential at the LHC: for this reason \muB\ is fixed to 1 MeV in the fit. $\phi$~and $K^{*0}$ are not included in the fit.

%\begin{figure}[htbp]
%\centering
%\includegraphics[width=0.4\textwidth]{2012-Aug-09-yields2_s2760_0-20sqm_noLambdaBarNopred}
%\includegraphics[width=0.4\textwidth]{2012-Aug-09-ratios2pi_alice0-20sqm_rhic_separated}
%\caption{Left: integrated yields at mid-rapidity \dndy~in central (0-20\%) \PbPb~collisions at \snn=2.76 TeV with results from thermal fit. Right: integrated yields at mid-rapidity \dndy~in central (0-20\%) \PbPb~collisions relative to pions with RHIC comparison and thermal model predictions.}
%  \label{fig:yields}
%\end{figure}
The temperature extracted from the fit \Tch\ = 152 $\pm$ 3 MeV is lower with respect to the temperature one would expect considering \Tch\ constant above SPS energies (164 MeV). From Figure \ref{fig:yields} (left) it is possible to see some tension between the data and the fit especially for strange and multi-strange particles. This is reflected also in the large value of the $\chi_2/N_{d.f.}$ = 39.6/9.
%The integrated yields relative to pions are reported in Figure \ref{fig:yields} (right), together with results from previous experiments at RHIC.
The comparison of integrated yields relative to pions (Figure \ref{fig:yields}, right) with RHIC hints to decreasing ratios at the LHC, especially for what concerns p/$\pi$ and $\Lambda$/$\pi$.
The prediction from the thermal model is reported with two different values of the freezeout temperature: \Tch\ = 164 MeV (value obtained from fit to RHIC data) and \Tch\ = 152 MeV (from the fit described above).
\Tch\ = 164 MeV seems to reproduce the multi-strange ratios quite well, but some discrepancy is observed for p/$\pi$ and $\Lambda$/$\pi$. On the other hand the model prediction using \Tch\ = 152 MeV obtained from the fit is closer to the measured p/$\pi$ and $\Lambda$/$\pi$ ratios but it misses ratios involving multi-strange hadrons.
It has already been pointed out that interactions in the hadronic phase, in particular via the large cross section channel for antibaryon-baryon annihilation, could explain the significant deviation from the usual thermal ratios \cite{Steinheimer2012,Becattini2012}.

\section{Conclusions}
\pt\ distributions of $\pi$, K, p in central (0-5\%) \PbPb~collisions at the LHC are harder than previously measured at RHIC. They are well described by hydrodynamic models including a refined description of the late fireball stages. Fitting the spectra with a hydrodynamic-inspired blast wave model results in the highest radial flow parameter ever measured, \avbT\ = 0.65 $\pm$ 0.02.

Event shape engineering is a powerful tool to select events with different values of the elliptic flow based on the magnitude of the flow vector $\vec{Q_2}$. 
The \pt\ distributions of $\pi$, K, p in semi-central (30-40\%) \PbPb~collisions show a correlation between radial flow and elliptic flow: high (low) elliptic flow events mean harder (softer) spectra.

In central (0-20\%) \PbPb~collisions at \snn\ = 2.76 TeV a significant deviation from the usual thermal production is observed, especially for p and $\Lambda$. The temperature obtained from thermal fit to integrated yields at mid-rapidity \dndy\ is lower than expected from previous experiments. This discrepancy can be explained in terms of interaction in the hadronic phase which can modify the relative hadron abundances. It should be noted that a refined description of the hadronic phase is also needed to reproduce femtoscopy correlations at the LHC \cite{MaciejQM2012}.

Identified hadron results at the LHC cast a new light upon the
hydrodynamic and thermal behavior of the hadron production in
heavy-ion collisions. The p-A run expected at the LHC at the beginning
of 2013, together with the continuously improving experimental
precision and description from the theory,
will provide further insights (and model constraints) on the heavy-ion puzzle.

%\bibliographystyle{model1-num-names}
%\bibliographystyle{plain}
%\bibliography{LM_QM2012_Proceedings}

\end{document}